# Development of a Static Magnetic Diode: The "Magnode" - The DC Case

G. Hathaway[1*]

[1]Hathaway Research International Inc., 1080 19th Sideroad, King City, Ontario, Canada

This report experimentally demonstrates the feasibility of applying an electrical current though a stationary graphene structure to induce magnetic non-reciprocity or, in other words, to create a static magnetic diode, and thereby validate the suggestion of Prat-Camps [1] et al. regarding the use of graphene for such a purpose. The described graphene-based static magnetic diode is assessed against the rotating copper magnetic diode of Prat-Camps and found to be comparable.

## I. INTRODUCTION

Magnetic non-reciprocity was discovered by Prat-Camps et al. [1] based on the movement of conduction electrons near a DC or low-frequency magnetic dipole electromagnet. The normal dipolar magnetic field of the dipole was "swept" along in one direction due to the action of the nearby moving conduction electrons. Thus, normal magnetic reciprocity (equal and opposite magnetic flux on either pole of a magnetic dipole) was broken for the first time, resulting in a "magnetic diode". To quote from [1]:

> "The use of a moving conductive material to break magnetic reciprocity boils down to the Lorentz force that the free electrons of the [moving] conductor experience as they move through the magnetic field." [ref 1 pg. 4]

The net result was the ability to direct a substantial amount of magnetic dipole flux in one direction while minimizing it in the opposite direction. Apparently no field strength was sacrificed in the operation of the Prat-Camps device. Other methods have been devised to minimize flux in one direction, but they typically require an absorptive member on one side of the magnetic dipole, thus resulting in the loss of approximately half the total available flux from the dipole.

As noted, the Prat-Camps device is based on a dipole being in close proximity to a moving flux of conduction electrons. This electron flux is provided by placing the magnetic dipole inside a U-shaped groove cut into a rotating copper disc, wherein the sides and bottom of this rotating copper groove provide the nearby conduction electron flux. Toward the end of the Prat-Camps article [1], the authors state:

> "In principle, one could replace the mechanical movement of the whole material by an externally applied electric field, which would force the electrons to move with a constant mean velocity in the conductor" [ref 1 pg. 4]

In addition:

> "Interestingly, other materials like graphene exhibit carrier mobilities that can be more than 3 orders of magnitude larger than in copper [2] while being able to sustain current densities on the order of $\sim 10^8$ A/cm$^2$ [3]. Hence, graphene is an interesting candidate to explore implementations that do not rely on mechanical movement of macroscopic objects." [ref 1 pg. 5]

In the first part of this paper experiments involving a rotating copper disk similar to those of Prat-Camps are described but using a permanent magnet dipole rather than a small electromagnet as the source of magnetic flux. The second part deals with investigations into the use of graphene to supply the moving conduction electrons in providing the nearby electric field necessary for magnetic diode non-reciprocity. The conduction electrons will be supplied by a high-current DC power supply.

## II. EXPERIMENTS INCORPORATING A PERMANENT MAGNET AND ROTATING COPPER CONDUCTOR

In order to duplicate the results of Prat-Camps, a variable speed DC motor was used. This motor drives a heavy copper ring into which circumferential and side face grooves are machined. The details are shown in Figure 1.

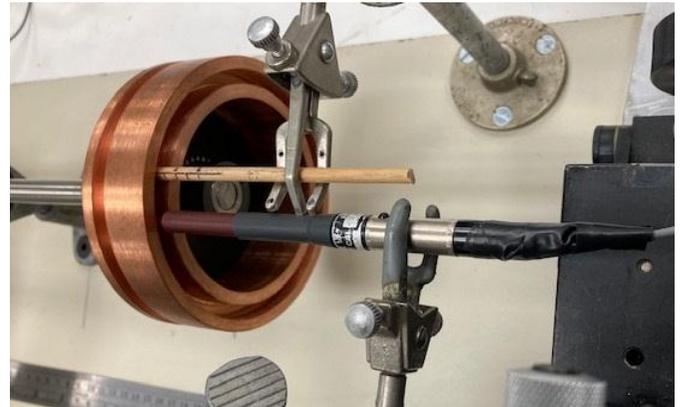

FIG 1. Detail of spinning copper conductor experiments.

Figure 1 shows the side groove experiment which is similar to that of Prat-Camps. The copper ring has an outer diameter of 114 mm and the side groove is 5.9 mm wide by 13.7 mm deep and starts 5.3 mm from the outer diameter of the disk. Thus, the outer copper width of the side groove wall is 5.3 mm and the inner copper wall thickness is 5.5 mm. The thickness of the floor of the groove is approx. 6 mm. The disk can spin up to approx. 1,500 rpm (25Hz) CW and CCW (+/- 10 rpm). These groove dimensions are somewhat smaller than those employed by Prat-Camps.

Figure 1 shows a wooden stick supported by a clamp onto the end of which a small permanent neodymium disk magnet of 5 mm diameter x 2.5 mm thick is glued, magnetized through the thickness. In this experiment, the magnet is inserted 12 mm into the side groove. The dipole moment, i.e., the axial direction of the magnet's flux, is perpendicular to the disk radius, i.e., is circumferential.

Inserted approx. 17 mm circumferentially away from the magnet is a transverse magnetic field Hall probe (Bell T-4048-001), whose active area is also centered at 12 mm into the side groove. The active area of the transverse probe is also perpendicular to the disk radius and thus in line with the direction of flux emanating from the magnet. This probe is connected to a gaussmeter (Bell model 4048). The gauss readings are +/-0.4 gauss.

A. Side groove tests

A typical example of the side groove measurements is shown in Fig. 2 for CCW (while facing the groove) rotation, i.e., for the spin direction from the magnet to the probe.

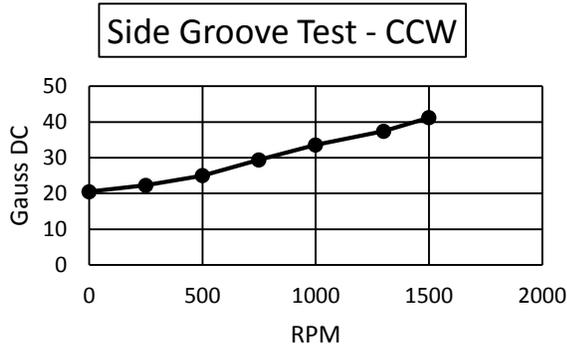

FIG. 2. Gauss vs. RPM for typical side groove test on Cu disk – rotation toward probe.

Here, an approx. doubling of the flux density is observed when the copper disk is spun to 1500 rpm.

Figure 3 shows the situation for CW rotation, i.e., away from the probe.

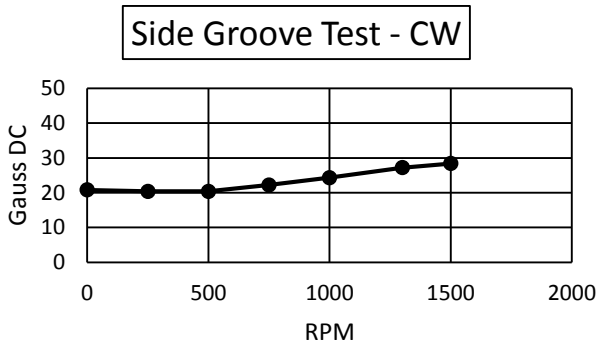

FIG. 3. Gauss vs. RPM for typical side groove test on Cu disk – rotation away from probe.

There is an obvious greatly reduced increase in flux density with rotational speed. Less obvious is why there is any increase in flux density with rotational speed at all. One answer may be gleaned from noting Prat-Camps [1] Figure 3 (b), which depicts the flux pattern calculated for a radially-oriented dipole under similar conditions as above. For "negative" rotation, i.e., CW, there is still magnetic flux coupling the two dipoles, the amplitude of which depends on the geometrical distortion of the field, which in turn depends on the rotational speed. This is more clearly shown in Figure 3 (c) in their paper, which shows increasing "negative" flux density with rotational speed in the CW direction that is not as pronounced as in the CCW direction.

In fact, the Prat-Camps ratio of average CCW to average CW flux density is approx. 2:1, similar to our results.

Finally, a test was performed where the magnet is rotated such that its axial flux is parallel to the disk radius and, thus, perpendicular to the probe in the CCW case – Fig. 4.

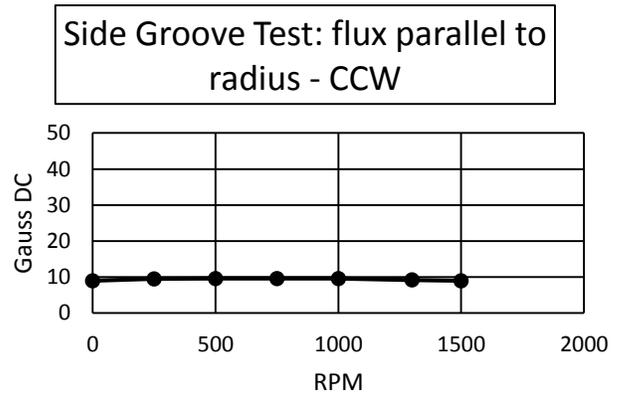

FIG. 4. Gauss vs. RPM for side groove with flux parallel to radius – CCW rotation.

Figure 4 demonstrates that the side flux of the magnet is only slightly altered by the disk rotation, as expected.

B. Top groove tests

Similar tests to the above were performed using a permanent magnet of the same size mounted onto a wooden stick and a transverse Hall probe both inserted radially into the groove on the top/outer surface of the copper disk. The top groove dimensions are 5.9 mm wide x 11.8 mm deep starting 5.3 mm from the left edge of the disk. The remaining flat surface to the right of the groove is 19.5 mm wide and was used for the flat surface tests in the following section. The transverse probe active area faced the axial flux direction of the magnet (flux direction perpendicular to disk radius), and both were separated by 18 mm. Figure 5 shows the results.

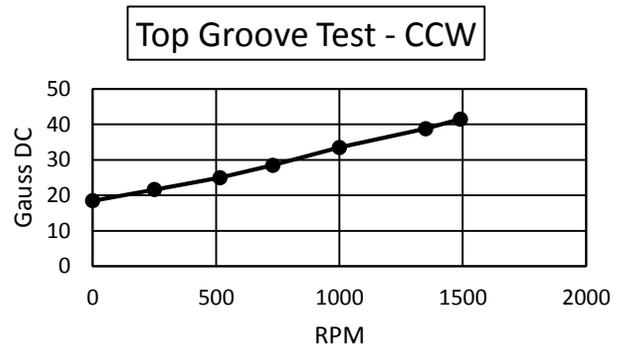

FIG. 5. Gauss vs. RPM – Top groove test – CCW rotation.

These results are substantially equivalent to the side groove tests, indicating that orientation of the magnetic field relative to the rotational axis of the rotating conductive body (copper disk) is immaterial. As long as there are moving conduction electrons nearby, the magnetic field will be altered.

### C. Top surface tests

The following test was performed to estimate the degree to which the mass of nearby rotating conductor, and thus the density of nearby conduction electrons, influences the distortion of the magnetic field. Here, the probe and magnet skim the top surface of the copper disk in the center of the 19.5 mm wide flat surface to the right of the top groove. Figure 6 shows the setup.

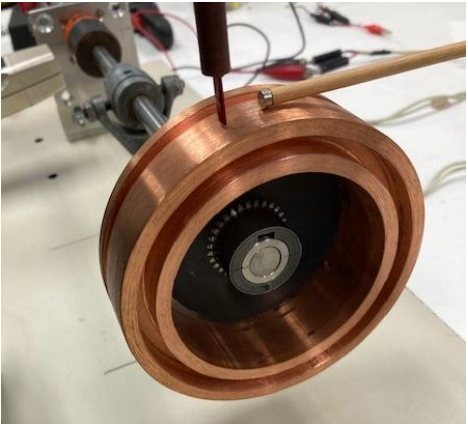

FIG. 6. Top surface experimental setup.

The probe and magnet are approx. 18 mm apart. Figure 7 shows the results.

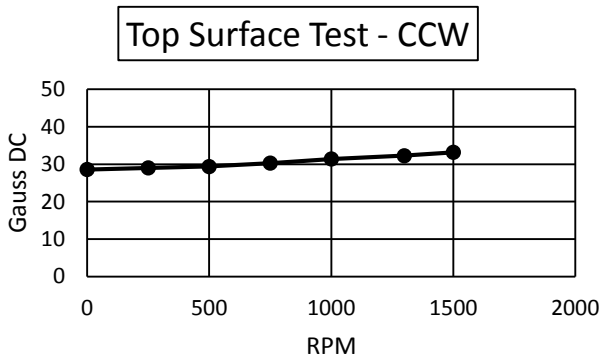

FIG. 7. Gauss vs. RPM for top surface tests – CCW.

There is clearly a small flux increase with rotational speed of approx. 0.16. The side groove flux improvement seen in Fig. 2. corresponds to an increase of approx. 1.0 times, with the difference being approx. 6 times enhanced due to the increased mass of nearby conductor. The degree to which the mass of nearby rotating conductors is increased in Fig. 1 vs. Fig. 6 can be very roughly estimated by comparing the copper volume of the surrounding groove in Fig. 1 (~ 240 mm$^3$) with the approximate volume to which the magnet is exposed in the top surface in Fig. 6 (~35 mm$^3$), resulting in a ~7 times volume (mass) enhancement. It therefore appears that the degree of magnetic field enhancement is roughly dependent on the mass of the nearby rotating conductor material.

Both the side groove and top groove rotating experiments showed there was an approx. 2 times increase in flux density in the direction of motion. In the experiments of Prat-Camps [1], an increased flux density (proportional to voltage across the pickup coil) of approximately the same magnitude was noted. These simple rotational tests confirm the observations of Prat-Camps, that magnetic reciprocity can be broken by the presence of moving conduction electrons.

### III. EXPERIMENTS INCORPORATING A PERMANENT MAGNET AND NO MOVING CONDUCTOR

The objective of this phase of the investigation is to obviate the use of a moving conductor by replacing it with a static device through which electrons can freely flow, according to the suggestion of Prat-Camps. In addition, the source of the magnetic field, i.e., the magnetic dipole, would ideally be completely surrounded by the moving electron flux, rather than on only three sides as in the rotating copper disc experiment of Prat-Camps. This more ideal construction will be described in a later section.

In the experiments below, the current was +/-1 ADC and the flux density +/-0.3 Gauss.

### A. Slotted copper bar tests.

Figure 8 shows the setup for the copper slot tests.

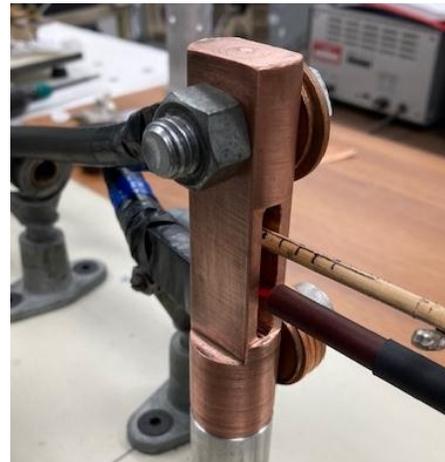

FIG. 8. The copper slot test setup in detail.

A 40 mm long, 6.3 mm wide and 13.8 mm deep slot was milled into a copper bar, leaving 4.2 mm and 4.4 mm thick side walls. DC current of up to 300 A flowed through the bar. Again, a small 5 mm diameter x 2.5 mm thick neodymium permanent magnet, magnetized through the thickness (flux in axial direction), was glued to one end of a wooden stick. The results are shown in Fig. 9 for the situation where the active surface of the probe is perpendicular to the axis of the magnet (flat faces parallel).

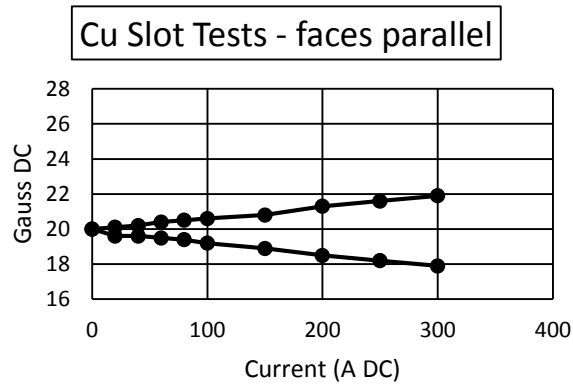

FIG. 9. Cu slot test – flat faces of magnet and probe parallel.

The upper of the two traces in Fig. 9 depicts current in one direction, and the lower trace depicts the flux density for current in the opposite direction. Here, there is obvious essentially linear enhancement, as well as an equal decrease in the flux density, whose direction depends on the direction of the current, although there is a significantly lower degree of enhancement than for a rotating conductor with comparable nearby conduction mass. Compared to the rotating situation, however, the degree of flux enhancement is symmetrical with the current direction. Assuming a linear extrapolation, it would take approx. 3000 A to double the flux density using the static slotted copper bar to obtain results equivalent to those of the rotating grooved copper disk.

### B. Graphene sheet experiments.

Several experiments using commercial graphene sheets (Graphene Supermarket, item 8x4-5P), each 25 μ in thickness, were undertaken in order to compare graphene with copper as a source of nearby moving conduction electrons. First, it was necessary to determine the ability of these thin graphene sheets to thermally withstand the high currents to which they would be subjected. The thermal measurements of layered sheets of graphene indicated that intermittent DC currents of maximum 100 A would be possible. Therefore, in the following graphene experiments, the current is limited to 100 A intermittent and extrapolated as necessary.

#### 1. Graphene slot experiments

An experiment was conducted in which the slotted copper bar was replaced by a U-shaped slot with graphene walls whose slot dimensions are comparable to those of the slotted copper bar. Figure 10 (a) shows details of the construction. Ten sheets of commercial graphene were cut in such a way as to provide walls and a floor for the slot as well as overhanging tabs for the introduction of high current. As shown in Fig. 10 (a), a wooden form is used to position the graphene sheets and hold them in place while the overhanging graphene tabs are folded against an inner brass plate. Once complete, an additional outer brass plate is screwed onto the inner plate to sandwich the graphene sheets and thus provide a high-current and mechanically robust connection to the high current power supply cables, as seen in Fig. 10 (b).

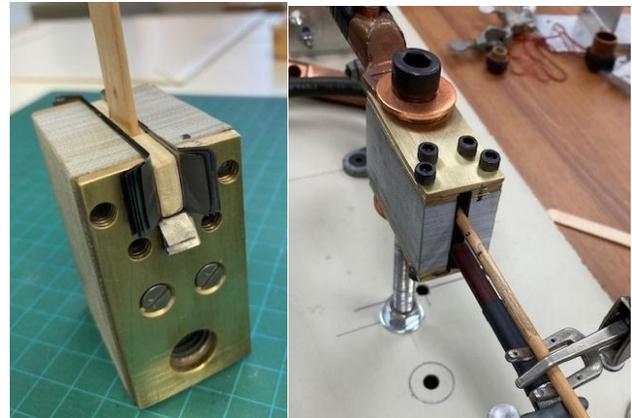

FIG. 10. (a) Graphene slot construction in detail.
FIG. 10. (b) Graphene slot experiment.

The dimensions of the graphene slot are 5.6 mm wide x 14.0 mm deep x 45 mm long. Figure 10 (b) also shows the position of the permanent magnet and transverse probe, which are 16 mm apart from one another with flat faces facing each other, as seen in Fig. 8.

The results are shown in Fig. 11 along with linear extrapolation to 300 A.

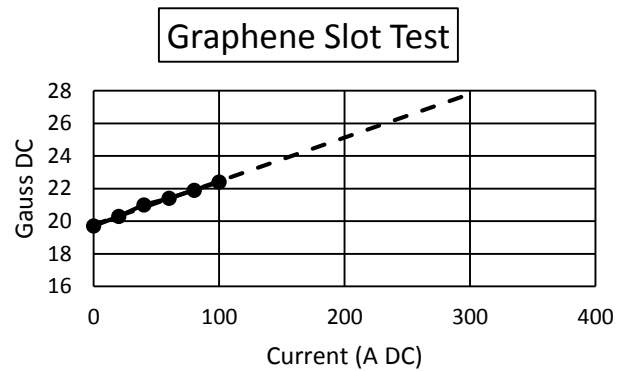

FIG. 11. Graphene slot experiment.

Recall that the copper slot test showed a 9.5% flux enhancement at 300 A (22 G vs. 20 G), whereas the graphene slot tests (above) show an approx. 41% (28 G extrapolated vs. 20 G) enhancement. A small part of this difference is due to the width of the copper slot being 6.3 mm vs. 5.6 mm for the graphene slot. Recall that the percent enhancement in the rotating copper slot experiments was about 100% (40 G vs. 20 G). Extrapolating the results in Fig. 11, doubling of the zero current flux density (100% enhancement) would require about 750 A for the 10 sheet graphene slot.

### C. Tubular Graphene "Magnode" Experiments

The above graphene slot experiments were performed in a slot with one open face, albeit with the open face being quite far from the magnet and probe. Therefore, a tubular cavity composed of graphene was chosen as the source of moving conduction electrons. The tube used for this experiment is a 20 cm long x 62 mm wide sheet of graphene foil of the type used in the graphene slot experiments. The sheet was rolled upon a phenolic tube of dimensions 6.3 mm ID x 7.9 mm OD x 45.4 mm long, resulting in an 8-layered tube. The extended ends were sliced and separated and then folded over an outer phenolic tube for support. Two brass plates were machined to clamp onto the sliced ends of the tube with allowances made for connection to high-current cables – see Fig. 12.

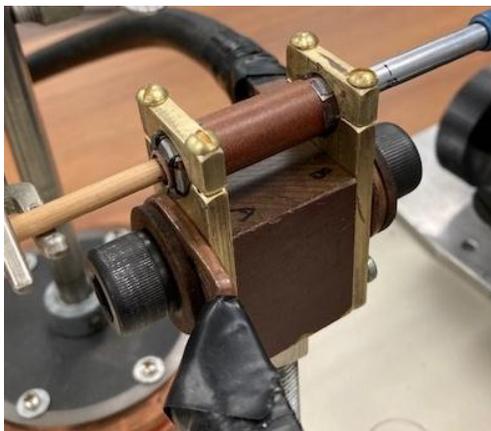

FIG. 12. Details of graphene tube "Magnode" mounting: magnet inserted on left, probe on right.

Figure 12 shows the method for connecting the high current cables to the graphene tube and the wooden dowel inserted into the left end of the tube, which has a 5 mm diameter x 2.5 mm thickness magnet glued to its end, as in Fig. 6. An axial gaussmeter probe (Bell HAB-92-2502) is inserted into the opposite end of the tube, which is connected to a Bell gaussmeter (Bell 9200).

Figure 13 shows a plot of the results. Currents were limited to 100 A to avoid overheating and linearly extrapolated to 300 A to allow comparison with previous results.

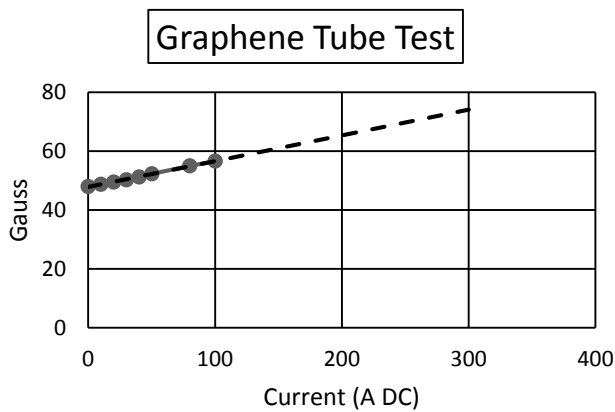

FIG. 13. Graphene tube "Magnode" experiment.

Using the enhancement concept highlighted above, up to 300 A the enhancement factor was 55% (74 G extrapolated vs. 48 G) vs. 41% for the graphene slot setup, clearly demonstrating the benefit of completely enclosing the source of the magnetic field with moving conduction electrons. Note that the top groove results for the rotating copper disk (Fig. 5) at 730 rpm result in an enhancement of ~55%. In addition, the graphene tube Magnode incorporated 8 wraps of graphene, whereas 10 sheets were used for the graphene slot experiment. Thus, a tubular setup likely results in even greater enhancement than that of a slot. However, for practical purposes, such as enhancing the magnetic field on one side of an electrical coil, it may be necessary to employ slot geometry.

IV. CONCLUSIONS

Moving conduction electrons in commercial-grade graphene sheets conducting DC current in either a static slot-type geometry or fully-enclosed static geometry can rival rotating copper conduction electrons in enhancing the asymmetry of an enclosed magnetic dipole, thus circumventing magnetic reciprocity.

As noted by Prat-Camps [1], graphene-based static magnetic diodes are expected to have many applications, including in wireless power transfer and transformers. Additional applications include electromotive devices and medical applications involving focused magnetic fields.

*george@hathawayresearch.com